# The $^{13}$CO-rich atmosphere of a young accreting super-Jupiter

Yapeng Zhang[1], Ignas A. G. Snellen[1], Alexander J. Bohn[1], Paul Mollière[2], Christian Ginski[3], H. Jens Hoeijmakers[4,5], Matthew A. Kenworthy[1], Eric E. Mamajek[6,7], Tiffany Meshkat[8], Maddalena Reggiani[9], Frans Snik[1]

[1]*Leiden Observatory, Leiden University, Postbus 9513, 2300RA Leiden, The Netherlands*
[2]*Max-Planck-Institut für Astronomie, Königstuhl 17, 69117 Heidelberg, Germany*
[3]*Anton Pannekoek Institute for Astronomy, University of Amsterdam, Science Park 904, 1098XH Amsterdam, The Netherlands*
[4]*Observatoire de Genève, Université de Genève, 51 Chemin des Maillettes, 1290 Sauverny, Switzerland*
[5]*Lund Observatory, Department of Astronomy and Theoretical Physics, Lunds Universitet, Sölvegatan 27, 223 62 Lund, Sweden*
[6]*Jet Propulsion Laboratory, California Institute of Technology, 4800 Oak Grove Drive, M/S 321-100, Pasadena CA 91109, USA*
[7]*Department of Physics & Astronomy, University of Rochester, Rochester NY 14627, USA*
[8]*IPAC, California Institute of Technology, M/C 100-22, 1200 East California Boulevard, Pasadena CA 91125, USA*
[9]*Institute of Astronomy, KU Leuven, Celestijnenlaan 200D, B-3001 Leuven, Belgium*

**Isotope abundance ratios play an important role in astronomy and planetary sciences, providing insights in the origin and evolution of the Solar System, interstellar chemistry, and stellar nucleosynthesis[1,2]. In contrast to deuterium/hydrogen ratios, carbon isotope ratios are found to be roughly constant (~89) in the Solar System[1,3], but do vary on galactic scales with $^{12}$C/$^{13}$C~68 in the current local interstellar medium[4-6]. In molecular clouds and protoplanetary disks, $^{12}$CO/$^{13}$CO isotopologue ratios can be altered by ice and gas partitioning[7], low-temperature isotopic ion exchange reactions[8], and isotope-selective photodissociation[9]. Here we report on the detection of $^{13}$CO in the atmosphere of the young, accreting giant planet TYC 8998-760-1 b at a statistical significance of > 6σ. Marginalizing over the planet's atmospheric temperature structure, chemical composition, and spectral calibration uncertainties, suggests a $^{12}$CO/$^{13}$CO ratio of $31^{+17}_{-10}$(90% confidence), a significant enrichment in $^{13}$C with respect to the terrestrial standard and the local interstellar value. Since the current location of TYC 8998 b at ≥160 au is far beyond the CO snowline, we postulate that it accreted a significant fraction of its carbon from ices enriched in $^{13}$C through fractionation. Future isotopologue measurements in exoplanet atmospheres can provide unique constraints on where, when and how planets are formed.**

TYC 8998-760-1 b[10] is a widely-separated planetary mass companion around a young solar analog TYC 8998-760-1 (also known as 2MASS J13251211-6456207) with an age of ~17 Myr[11]. With the recent detection of a second planet[12], it is part of the first directly imaged multi-planet system around a solar-type star. TYC 8998 b is located at a projected separation of 160 au, with an estimated mass of 14±3 $M_J$. We observed the planet on two nights, 2019 June 5 and June 19, using Spectrograph for INtegral Field Observations in the Near Infrared (SINFONI)[13,14] installed at the Cassegrain focus of UT3 of the Very Large Telescope of the European Southern Observatory at Cerro Paranal, Chile. The observations were performed in K-band (1.95 - 2.45 μm), providing a spectral resolving power (λ/Δλ) of ~4500. We extracted the spectrum of TYC 8998 b from 2.10 to 2.45 μm as detailed in Methods.

As shown in Figure 1, the planet spectrum is dominated by molecular features from $H_2O$ and CO. The $^{12}$CO $v$=2-0, 3-1, 4-2 bandheads are visible at 2.2935, 2.3227, and 2.3535 μm respectively. When we compare the observed spectrum with the best-fit model obtained by atmospheric retrieval, an extra emission signature at 2.166 μm is seen in Figure 1c, which is identified as the hydrogen Brackett γ recombination line. This is likely an indication of on-going accretion of circumplanetary material onto the planet. We estimated a mass accretion rate of $10^{-9.4 \pm 1.3}$ $M_\odot$ yr$^{-1}$ using the Br γ line luminosity (see Methods). Future observations at longer wavelengths and polarimetric data can provide further insights into the circumplanetary disk and accretion process[15].

In order to characterize the atmosphere of the planet, we performed a Bayesian retrieval analysis on the spectrum using the radiative transfer tool petitRADTRANS (pRT)[16], connected to the nested sampling tool PyMultiNest[17]. We focus on revealing the presence of $^{13}$CO, which is expected to be the most detectable isotopologue in atmospheres of gas giants[18], and measuring the isotopologue abundance ratio $^{12}$CO/ $^{13}$CO. In addition, we aim to constrain atmospheric properties of the planet such as the carbon-to-oxygen ratio C/O, which may shed light on the conditions during the formation of the planet[19].

We set up the retrieval model in a similar way as used in previous work on HR 8799 e[20] (see Methods). The model consists of nine free parameters: the planet radius $R_P$, surface gravity log($g$), metallicity [Fe/H], carbon-to-oxygen ratio C/O, the isotopologue abundance ratio $^{12}$CO/ $^{13}$CO, three parameters for the Temperature-Pressure (T-P) profile, and a spectral slope correction factor $f_{slope}$, arising from uncertainties in the calibration of the observed spectrum. We ran retrievals on the spectrum of TYC 8998 b using two sets of models: the full model including $^{13}$CO and the reduced model excluding $^{13}$CO. The setups of the two models are identical except that we removed $^{13}$CO from the opacity sources (and therefore the $^{12}$CO/ $^{13}$CO parameter) in the latter. The best-fit model spectra are compared to the observed spectrum shown in Figure 1. The difference between the two models is visible at the $^{13}$CO bandheads around 2.345 μm (see Figure 1b). The full model provides a significantly better fit in terms of $\chi^2$ (see Extended Data Figure 1c) as a result of the additional $^{13}$CO opacity.

To qualitatively reveal the signature, we compare the observational residuals (that is, the observed spectrum with the best-fit reduced model subtracted off) to the noise-free template of the $^{13}$CO signal (that is, the best-fit full model subtracted by the same full model with the $^{13}$CO abundance set to zero). The observational residuals follow the expected $^{13}$CO signal, especially the dip caused by the $^{13}$CO bandheads and well-aligned individual lines (Figure 1d), which indicates the presence of $^{13}$CO in the data. We also cross-correlated the residuals with the $^{13}$CO model (Figure 1e). The cross-correlation function (CCF) between the observation and model follows the auto-correlation (ACF) of the model well. The broad feature in the CCF (and ACF) reflects the effect of $^{13}$CO bandheads, and the peak at zero radial velocity co-adds the individual lines of the $^{13}$CO signal.

In terms of the significance of the $^{13}$CO detection, comparing the Bayesian evidence ($Z$) of the reduced model with the full model, allows us to assess the extent to which the model including $^{13}$CO is favored by the observations. In Bayesian model comparison, the Bayes factor $B_m$ (calculated by the ratio of $Z$) is used as a proxy for the posterior odds ratio between two alternative models[21]. As a result, the Bayes factor between two models is $\ln(B_m) = \Delta\ln(Z) = 18$, meaning that the observation favors the full model (including $^{13}$CO) at a significance level of > 6σ.

The central values of the inferred parameters and their 90% uncertainties from the atmospheric retrieval are $R_P = 1.82 \pm 0.08$ $R_J$, $\log(g) = 4.51^{+0.34}_{-0.29}$, [Fe/H] = $0.07^{+0.31}_{-0.18}$, C/O = $0.52^{+0.04}_{-0.03}$, and $^{12}$CO/$^{13}$CO = $31^{+17}_{-10}$ (Figure 2b and 2c). Hence, while the C/O ratio is measured to be near the solar value, the planet atmosphere is observed to be rich in $^{13}$CO. The inferred T-P profile and posterior distribution of parameters for both models are shown in Extended Data Figure 2.

The $^{12}$CO/$^{13}$CO ratio of $31^{+17}_{-10}$ we infer for TYC 8998 b is lower (2.5σ) than measured in the local interstellar medium (ISM) ~68[4,5], with the latter also lower than the ratio of ~89 observed in the Solar System[1], which is partly thought to be the consequence of galactic chemical evolution, and reflects the relative degree of primary to secondary processing in stars[4-6]. Young systems formed in the local environment are expected to inherit the $^{12}$C/$^{13}$C ratio of ~68. However, the measurements of complex carbon-bearing molecules (which formed at low temperature from CO on grains) towards protostars result in a low $^{12}$C/$^{13}$C ratio of ~30, indicating an enhancement of $^{13}$CO or $^{13}$C in the ice[22,23]. This may be attributed to carbon fractionation processes, including isotopic ion exchange reactions[8], isotope-selective photodissociation[9] and ice/gas isotopologue partitioning[7]. The isotopic ion exchange reactions enhance the $^{13}$CO abundance in the gas at low temperatures, followed by the freeze-out of CO gas onto the grains to increase the $^{13}$CO abundance in the ice[22]. The ice/gas isotopologue partitioning, due to slightly different binding energies of the two isotopologues, may contribute to the enrichment of $^{13}$CO in the ice, but only in very narrow temperature ranges as shown by laboratory experiments[24]. The isotope-selective photodissociation alters the $^{12}$CO/$^{13}$CO ratios at different layers in circumstellar disks because the self-shielding of the rarer isotopologues kicks in at a deeper layer into the disk[25]. The preferential photodissociation of $^{13}$CO generates more atomic $^{13}$C which freezes out onto the ices in the midplane to enrich the $^{13}$C in other complex carbon-bearing molecules. Detailed modeling incorporating these fractionation processes in protoplanetary disks suggests that the $^{13}$CO in the gas could also be enhanced at intermediate layers[25]. Planets may be able to accrete the $^{13}$CO-rich gas from these intermediate layers through vertical accretion[26]. Alternatively, with the warm gas cycling to cold regions through vertical mixing and then freezing out onto grains, the enrichment may be inherited by pebbles and planetesimals forming there.

In this light, we postulate a framework to explain the $^{13}$CO-rich atmosphere of TYC 8998 b, and the near-constant $^{13}$C abundance in solar-system objects. It is generally accepted that in the inner parts of a protoplanetary disk, carbon is mostly present in the gas phase as CO, but locked into ices at larger distances, in particular in the disk midplane (Figure 3). The CO-snowline, the transition region between the gaseous and solid phase, is governed by the level of stellar irradiation and is expected to be at ~20 au for young solar-mass stars[27] such as TYC 8998. The current location of TYC 8998 b at >160 au is so far out that it has likely formed outside the CO-snowline. This implies that it has accreted the bulk of its carbon from ices, which is generally enhanced in $^{13}$CO and $^{13}$C as is discussed above, resulting in the observed $^{13}$CO-rich atmosphere. Our spectral analysis also points to an atmospheric C/O ratio of $0.52^{+0.04}_{-0.03}$, further suggesting that the planet was indeed formed beyond the CO-snowline[19]. Similarly, for other widely separated exoplanets formed outside the CO-snowline with significant ice accretion, we would also expect $^{13}$CO-rich atmospheres.

Since, in contrast, the solar-system planets are thought to be formed within the CO-snowline, we argue that the lack of any substantial enrichment in $^{13}$C in their atmospheres is because the bulk of their carbon reservoirs originate from CO gas, not ices. Although Oort-cloud comets now reside in the outer Solar System far beyond the snowline, they may have also formed in the inner disk and later been scattered out through the interaction with giant planets[28], therefore exhibiting the same $^{13}$C ratios as other objects in the Solar System.

As an analogy of $^{13}$CO enrichment through ice accretion, the deuterium to hydrogen ratios (D/H) of Uranus and Neptune are found to be enhanced in comparison with Jupiter, attributed to an increasing contribution from accretion of HDO-rich ices beyond the water snowline[29]. With the caveat that carbon-fractionation is orders of magnitude smaller than D-fractionation, it could still result in a $^{13}$C enrichment of a factor of two. For a further understanding of the role of carbon-fractionation in planet formation it will be important to obtain a quantitative assessment of the effects through detailed disk modelling. Future measurements of exoplanet $^{12}$C/$^{13}$C (and D/H) ratios[18,30] can provide an exciting new way to constrain when, where, and how planets are formed.

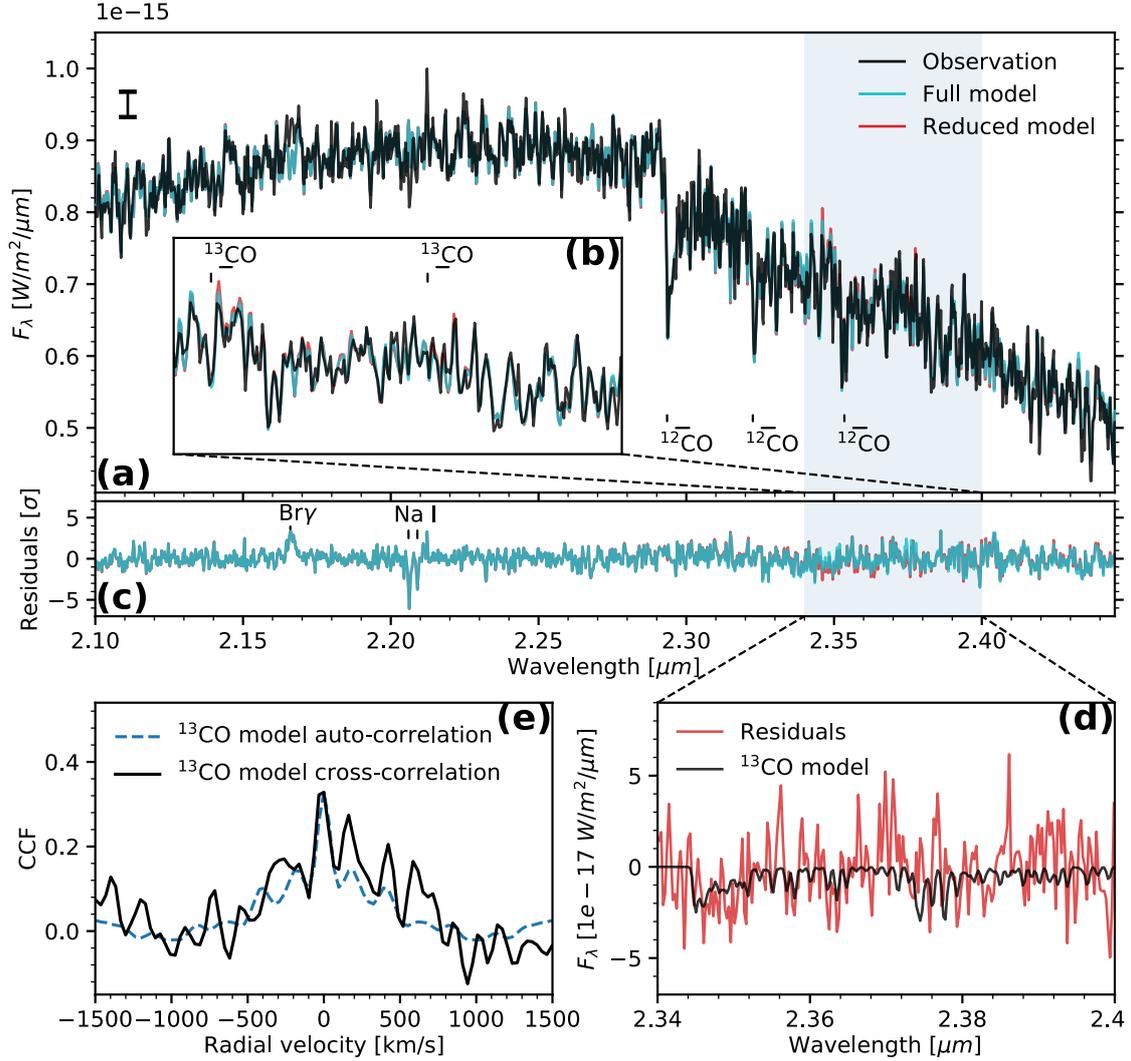

**Figure 1 | Observed SINFONI spectrum of exoplanet TYC 8998 b and cross-correlation signal of $^{13}$CO.** Panel (a): The observed spectrum is shown in black, the best-fit model including all carbon isotopes in cyan, and the best-fit reduced model, without $^{13}$CO, is shown in red. The typical uncertainty per pixel in the observed spectrum is denoted by the error bar (1σ) on the upper left. Sub-panel (b) shows a zoom-in over the 2.34 - 2.40 μm spectral region, centered on the $^{13}$CO bandhead. Specific opacity from $^{13}$CO is located where the reduced model without $^{13}$CO (in red) is higher than the observed spectrum. Panel (c) shows the residuals (spectrum minus best-fit models, with the residuals of the reduced model in red) revealing the Brackett γ emission feature at 2.166 μm, and Na I absorption lines. Panel (d): Observational residuals in red are compared with the $^{13}$CO absorption model in black (the difference between the best-fit full model minus the same model with the $^{13}$CO opacity removed), revealing the similarity between the observations and models. Panel (e): Cross-correlation functions (CCF) between observational residuals and the $^{13}$CO model, as shown in black. The dashed line shows the auto-correlation of the $^{13}$CO model, scaled to the peak of the CCF.

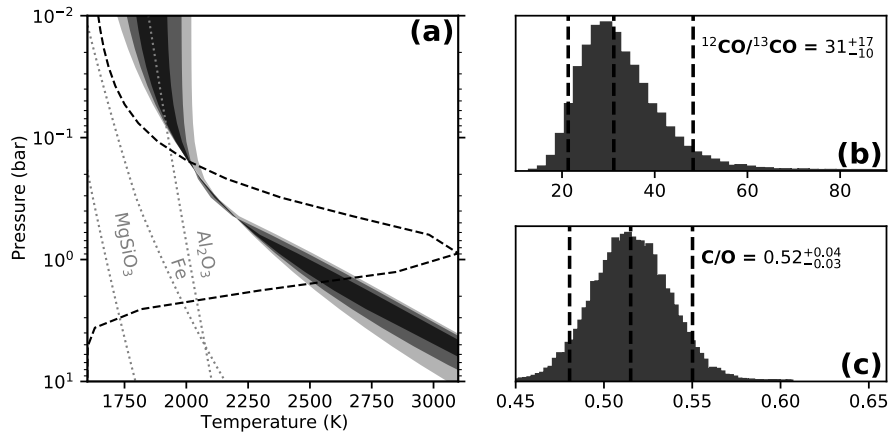

**Figure 2 | Spectral retrieval results.** (a) The retrieved temperature-pressure profile. The shaded regions with decreasing grey scale show 1σ, 2σ, and 3σ temperature uncertainty envelopes respectively. The gray dotted lines represent the condensation curves of three potential cloud species. The black dashed line shows the flux average of the emission contribution function. (b) Posterior distribution of CO isotopologue abundance ratio $^{12}$CO/$^{13}$CO. The vertical dashed lines denote the 5%, 50%, 95% quantiles (90% uncertainties) of the distribution. (c) Posterior distribution of carbon to oxygen ratio C/O.

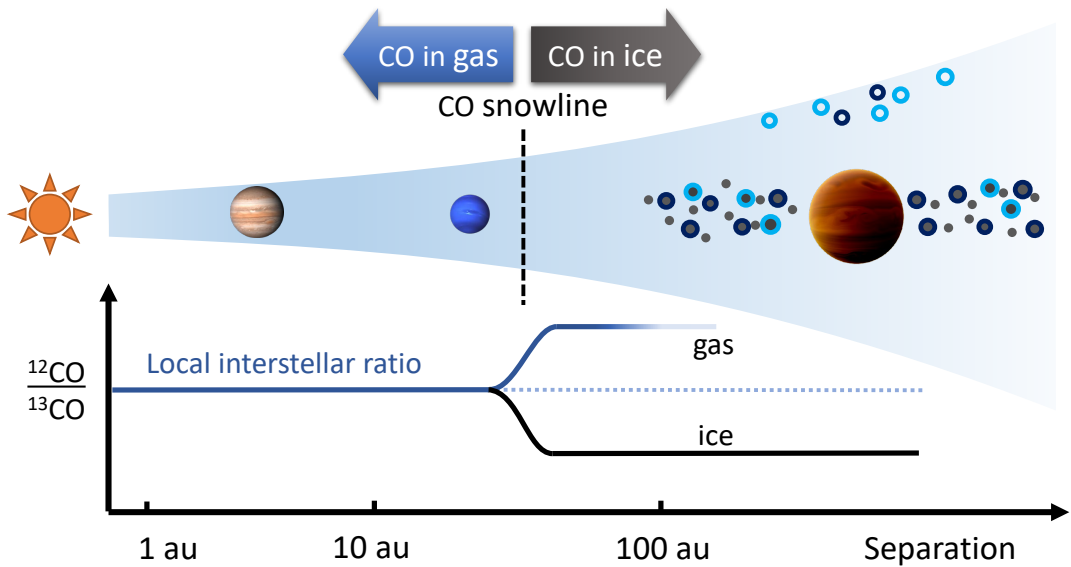

**Figure 3 | Cartoon of the birth environments of planets in a proto-planetary disk.** The two planets inside the CO snowline denote Jupiter and Neptune at their current locations, while TYC 8998 b is formed far outside this regime, where most carbon is expected to have been locked up in CO-ice and formed the main reservoir of carbon in the planet. We postulate that, this far outside the CO snowline, the ice was $^{13}$CO- or $^{13}$C- rich through carbon fractionation, resulting in the observed $^{13}$CO-rich atmosphere of the planet. A similar mechanism has been invoked to explain the trend in D/H within the Solar System. Future isotopologue measurements in exoplanet atmospheres can provide unique constraints on where, when and how planets are formed.

# Methods

**Observations and data analysis.** We observed TYC 8998 b on two nights, 2019 June 5 and June 19, using Spectrograph for INtegral Field Observations in the Near Infrared (SINFONI)[13,14] installed at the Cassegrain focus of UT3 of the Very Large Telescope of the European Southern Observatory at Cerro Paranal in Chile, under the ESO programme 2103.C-5012(C) (PI: Bohn). The observations were performed in K-band (1.95 - 2.45 μm), providing a spectral resolving power of ~4500. The spatial pixel scale is 0.025" per image slice, therefore 32 slices in total amount to a field of view (FOV) of 0.8". Both the primary star and the planet c are outside the FOV (see Extended Data Figure 1). As a result of the large planet-star separation (~1.7") and the planet-to-star contrast (Δmag~6.4)[10], the starlight contamination is small. We obtained 2 × 24 science frames with an NDIT of 2 and an exposure time of 150 seconds each. The observations were performed in pupil stabilized mode. During the first night, the airmass ranged from 1.31 to 1.34, the seeing varied from 0.81" to 2.63", and the atmospheric coherence time reached 12 milliseconds with the intervention by clouds. Given the unstable atmospheric conditions of night 1, a second night of observation was performed, delivering a seeing ranging from 0.51" to 1.07" at an airmass between 1.31 and 1.41 and a coherence time between 2.4 and 9.1 milliseconds. During each night, we took three frames of the sky before, amidst, and after the science frames with the same exposure setup. In addition to the science target, one featureless B-type standard star was observed each night, HIP 57861 and HIP 72332 respectively, serving for telluric and spectrophotometric calibration.

The raw data was first reduced using the SINFONI pipeline to correct for bias, flat fielding, sky background, and bad pixels. Then we obtained the intermediate 2D data products, which are composed of 32 vertical strips, each representing the spectra taken over an individual slitlet with its own wavelength solution. The target spectrum shows as vertical trails on slitlet images. For each science exposure, we extract the target spectrum from the intermediate 2D frame. We focus our analysis on the brightest four slitlet images, where the spectrum was extracted using the optimal extraction algorithm[31]. This algorithm accounts for background (including the starlight contamination) and automatically rejects outliers caused by bad pixels and cosmic rays, and delivers an optimal signal-to-noise ratio (SNR). We subsequently obtained four target spectra for every science exposure. All the spectra were then linearly interpolated to a common wavelength grid and optimally combined to form a combined spectrum for each night. During the process, we discarded two frames taken in night 1, of which the data quality was too low to clearly identify the target in the data (using a threshold of 3×background noise level) as a result of the poor atmospheric conditions.

The broad-band shapes of the final spectra per night obtained above require cautious calibration. First, the efficiency of the instrument+telescope varies over wavelength. This effect can be corrected by comparing the observed spectral shape of the standard star to its theoretical spectrum. For the standard stars in night 1 and 2, we used PHOENIX stellar models[32] with an effective temperature of 13500 K and 10800 K respectively[33,34]. Since the observed standard star spectra are firmly in the Rayleigh-Jeans regime, potential uncertainties in effective temperature have only marginal effects on the spectral slope. Second, atmospheric dispersion affects the overall spectral shape differently for different slitlets. The optimal extraction method we adopted can further aggravate this bias as different weights were assigned to slitlets when combining the spectra. To determine the most accurate spectral shape, we extracted spectra again while avoiding weighting. We started from the data cubes reconstructed by the SINFONI pipeline, and took a circular aperture of ∼5×FWHM to sum up the flux at each wavelength, ensuring that all the flux from the object was included. This plain extraction method was performed to the target as well as standard stars on both nights. In this way, we derived spectra for the target, which have lower SNR but a more accurate spectral shape. The broad-band shape of this spectrum for each night was then applied to the optimally extracted spectrum by fitting a second order polynomial to the division between both spectra. After the spectrophotometric correction, the shape of the combined spectra for the two nights differ by ∼2% between 2.1 and 2.4 μm.

We used the ESO sky software tool Molecfit[35] v3.0.1 to perform telluric corrections on the final spectra of each night. The tool uses a Line-By-Line Radiative Transfer Model (LBLRTM) to derive telluric atmospheric transmission spectra that can be fitted to observations. For telluric model fitting, we removed a preliminary planetary model from the observed spectrum to minimize effects from the planetary spectrum on the telluric features. The fitting wavelength region is 2.19 - 2.43 μm, which contains the majority of strong telluric lines caused by $CH_4$, and $H_2O$ in Earth's atmosphere. Molecfit accounted for molecular abundances, instrument resolution, continuum level and wavelength solution that can fit observations best. The atmospheric transmission model for the entire wavelength range was then derived based on the best-fit parameters. Subsequently, the telluric model was removed from the combined spectrum to obtain the telluric-corrected spectrum. However, the correction is not perfect, resulting in artifacts at the red end of the spectra. The wavelength region beyond 2.4 μm, not important for our goals, is therefore masked in further analysis.

After telluric removal, we combined spectra of both nights into one master spectrum. We noted that the spectrum of the first night turned out to be of lower signal to noise than that of the second night, as expected from the observational conditions. We shifted the two spectra to the planetary rest frame and weighted-combined them by signal-to-noise ratio (SNR) squared. The SNR of each spectrum was measured by the standard deviation of the residuals (i.e. the observation minus the model spectrum generated by retrieval fitting) in the wavelength range from 2.35 to 2.4 μm. The master spectrum was finally scaled to the photometric flux of 8.8 × $10^{-16}$ W/m²/μm at 2.25 μm[10]. In this way, we obtained a final master spectrum of the companion shown in Figure 1, with a SNR of ∼50 near 2.2 μm and ∼40 at the red end. We note that the uncertainty per wavelength-step is almost constant along the wavelengths and independent of the flux of the object because the observations are read-noise limited. The red end of the spectrum therefore has lower SNR because of the lower object flux level and strong telluric absorption.

**Atmospheric retrieval model.** For our atmospheric retrieval we performed a Bayesian analysis using the radiative transfer tool petitRADTRANS (pRT)[16], connected to the nested sampling tool PyMultiNest[17], which is a Python wrapper of the MultiNest method[36]. To model the temperature structure of the planet atmosphere, we consider two classes of parameterization of the Temperature-Pressure (T-P) profile: the analytical and flexible model. The first temperature model involves analytical solutions for self-luminous atmospheres assuming a gray opacity. We set the temperature according to Eddington approximation: $T(\tau)^4=0.75T_{int}^4(2/3+\tau)$, where the optical depth $\tau$ is linked to the pressure by $\tau=\delta P^\alpha$. We defined $P_{phot}$ as the pressure where $\tau=1$. Hence, $\delta=P_{phot}^{-\alpha}$. This temperature model therefore has three free parameters, $T_{int}$, $\alpha$ and $P_{phot}$. The Eddington solution leads to an isothermal upper atmosphere. Although this seems unrealistic, the model still works well because the medium-resolution observations barely probe that high up in the atmosphere. At low altitudes, the atmosphere transitions to convective, where we force the temperature gradient onto a moist adiabat.

In the flexible temperature model, we focus on the temperature from 0.01 bar to 10 bar, where the contribution function of the observed spectrum peaks. The temperature outside this range is considered to be isothermal. We set four temperature knots spaced evenly on a log scale pressure within 0.02 to 5 bar. The T-P profile is obtained by spline interpolation of the temperature knots in the log space of pressure. There is no physical reasoning behind this T-P profile, therefore imposing less prior constraints on the solution.

The chemistry model used in our retrievals is detailed in ref-20. In short, the chemical abundances are determined via interpolation in a chemical equilibrium table using pressure P, temperature T, carbon-to-oxygen ratio C/O, and metallicity [Fe/H] as inputs. Then pRT computes synthetic emission spectra using temperature, chemical abundances, and surface gravity as inputs.

We tested the retrieval framework to ensure parameters can be correctly recovered by using synthetic spectra. Using the forward model presented above, we generated a mock spectrum at 2.1 - 2.45 μm, with a spectral resolution of 4500. The flux error was set to be constant, leading to a SNR of around 50 at 2.2 μm, which is similar as in the master spectrum. The mock spectrum was then perturbed by random noise generated according to this uncertainty in flux. We were able to recover all parameters within 1σ interval, validating the capability of the retrieval model.

**Retrievals on the TYC 8998 b spectrum.** We set up the retrieval model using the analytical T-P profile, with eight free parameters, $R_P$, $\log(g)$, [Fe/H], C/O, $T_{int}$, $\alpha$, $P_{phot}$ and $^{12}CO/^{13}CO$. The priors of these free parameters are listed in Extended Data Table 1. In the full model, we included $^{12}CO$, $H_2O$, $CH_4$, $NH_3$ and the isotopologue $^{13}CO$ as line opacity species, and the collision induced absorption of $H_2$-$H_2$, $H_2$-He. As for the reduced model, we removed $^{13}CO$ from the opacity sources. We used the line-by-line mode of pRT to calculate the emission spectra at high spectral resolution. To speed up the calculation, we took every fifth point of opacity tables with $\lambda/\Delta\lambda\sim10^6$. This sampling procedure, which has been benchmarked against the full tables, shows negligible effects on the synthetic spectra and retrieval results. The synthetic high-resolution spectra were convolved with a Gaussian kernel to match the resolving power of the instrument ($\lambda/\Delta\lambda\sim4500$), then binned to the wavelength grid of the master spectrum, and scaled to the observed flux according to $R_P$ and distance of the target.

In addition, as we noted above, there exists up to 2% discrepancy in terms of the broad-band spectral shape between two nights of observations. Therefore, before comparing the synthetic model spectra to the observation, we multiplied the models by a linear slope spanning from 1-$f_{slope}$ to 1+$f_{slope}$ in the wavelength range from 2.1 to 2.4 μm, where we introduced the slope correction factor $f_{slope}$ as a nuisance parameter to marginalize the uncertainty in the spectral shape. The retrieval process was performed by PyMultiNest, which uses 4000 live points to sample the parameter space and derives the posterior abundances of the fit. We ran MultiNest in Importance Nested Sampling mode with a constant efficiency of 5%. The outcome is shown in Extended Data Figure 2. We also utilized the flexible T-P profile to consolidate the retrieval results. The inferred parameters are found to be robust for different T-P model setups.

The retrieved value of log($g$) should be treated with caution, because we note that any small changes in the wavelength coverage of the fitting or the broad-band shape of the spectrum can result in significantly different values. As also indicated in other retrieval studies[37], CO and $H_2O$ features at K-band are not very sensitive to log($g$). Free retrievals at K-band alone are therefore unable to place robust constraints on the surface gravity and hence the planet mass. We also run retrievals with prior constraint on the planet mass. The solution converges to a lower surface gravity and the same isotopologue ratio, suggesting that the inaccurate log($g$) does not affect the inference of the $^{12}CO/^{13}CO$ ratio.

We compared the Bayesian evidence ($Z$) of the reduced model with the full model (Extended Data Figure 1c), the difference of which is related to the Bayes factor $B_m$ that translates to a frequentist measure[21] of > 6σ significance of the $^{13}CO$ detection. Although the evidence is dependent on the prior of the $^{13}CO$ abundance (Extended Data Table 1), we noted that increasing the prior range of log($^{13}CO/^{12}CO$) from (-12,0) to (-20,0) did not lead to significant change in the final evidence, and both of the priors are much broader than the ratio that one would expect in reality.

**Effects of clouds.** Clouds are ubiquitous in planetary atmospheres and can play an important role in the interpretation of spectra. We investigated their potential effects on the $^{13}CO$ measurement by including clouds in the retrieval models. We used the cloud model detailed in ref-19, which introduced four additional free parameters: the vertical eddy diffusion coefficient $K_{zz}$, the settling parameter $f_{sed}$, the width of the log-normal particle size distribution $\sigma_g$ and the mass fraction of the cloud species at the cloud base log($X$). The temperature of TYC 8998 b is too high to form Mg and Fe bearing condensates in the atmosphere. We therefore considered a more refractory species $Al_2O_3$ as the source of cloud opacity[38]. After testing cloudy models, we found the solution converged to cloud-free atmospheres, and the inferred parameter values remain unaffected, because the atmospheric temperature (Figure 2a) is too warm to form optically thick clouds that could make a significant impact on the spectrum. Therefore, we did not include clouds in our nominal models.

**Constraining mass accretion rate.** The Brackett γ emission line (see Figure 1) provides constraints on the mass accretion rate of the planet. We measured a Br γ line flux of 7.6 (±0.9) × $10^{-18}$ W/m², and a line luminosity log($L_{Brγ}/L_\odot$) of -5.7±0.5, from which we estimated the accretion luminosity log($L_{acc}/L_\odot$) to be -2.7±1.3 using the linear correlation given by ref-39.

Combining this with the retrieved values of radius and surface gravity, we derived a mass accretion rate of $10^{-9.4\pm1.3}$ $M_\odot$ yr$^{-1}$.

## Additional assessments of the reliability of $^{13}$CO detection in TYC 8998 b

We further assess the reliability of the $^{13}$CO detection in several ways. We study the $^{13}$CO signal for the individual nights and that from the first and second bandheads separately, investigate the impact of inaccurate telluric line removal, and the effects of possible other opacity sources. In addition, we show that for an easier accessible, nearby, isolated brown dwarf, archival spectroscopy data of the same $^{13}$CO features at higher spectral resolving power, show a strong signal - as expected from our TYC 8988 b observations (albeit with a more typical $^{12}$CO/$^{13}$CO ratio).

**Analysis on data of individual nights and bandheads:** To assess whether the $^{13}$CO signal is present in both nights, we performed the retrieval and cross-correlations analyses as detailed above for each night separately. The retrieval results are shown in the Extended Data Figure 3. $^{13}$CO is detected in both nights but at a lower significance, as expected. The derived constraints on $^{12}$CO/$^{13}$CO are consistent with each other. There is a ~2% difference in the slope correction factor $f_{slope}$ in the two nights, as was already noted in the spectral calibration above. This likely contributes to the slight discrepancies in the derived values for log($g$) and C/O. The cross correlation functions are shown in Extended Data Figure 4, also indicating that $^{13}$CO is detected in both nights individually, albeit at lower significance.

We also investigated whether the two bands of $^{13}$CO, starting at 2.345 and 2.374 μm, can both be detected individually. We examined this by separating the wavelength region into two parts, 2.34-2.37 and 2.37-2.40 μm, then calculating the cross-correlation in each part. Prior to the cross-correlation, the residuals were high-pass filtered using a Gaussian kernel with a width of 3 nm to remove the low-frequency variation in residuals and enhance the signal. The low-frequency variation, especially apparent in the second part of the data, introduces broad-band shape in the CCF which suppresses the signal. The filtered residuals are shown in panel c of the Extended Data Figure 4, and the cross-correlation results in panel d. The CCF for both bands show peaks at zero velocity, while the signal from the second band is less significant. We argue that although the expected absorption from the two bands is similar, the signal-to-noise of the spectral data at the 2.374 μm bandhead is lower due to the stronger telluric absorption in this region. This is highlighted in panel a of Extended Data Figure 5, showing the telluric transmission spectrum, observational residuals from the reduced model (without $^{13}$CO), and the $^{13}$CO-model. We believe that it is for this reason that the $^{13}$CO signal from the second bandhead is not as significant as from the first bandhead.

**Impact of telluric absorption:** Panel b of Extended Data Figure 5 shows the cross-correlation function between the $^{13}$CO model and the telluric line spectrum, demonstrating no evident signal. It indicates that the observed $^{13}$CO signal is not caused by any under or overcorrection of telluric lines.

**Other potential opacity sources.** In the retrieval models, we include the major molecular opacity sources (CO, $H_2O$, $CH_4$, $NH_3$) that play a role at the K-band. We inspected other potential absorbers given the expected temperature, including $CO_2$, HCN, and $C_2H_2$. None of them show significant features at the same wavelength region as $^{13}$CO. Moreover, we performed retrieval analysis to marginalize over these absorbers. Adding these molecules to the retrieval model does not change the inferred value of any parameter. Therefore, it is not plausible that the detected $^{13}$CO feature is caused by these molecules.

**Archival reference data of the young brown dwarf 2M0355:** To boost confidence in the detection of $^{13}$CO in TYC 8998 b, we analysed archival data of the young brown dwarf 2M0355[40]. This object resembles in several ways a young super Jupiter like TYC 8998 b, but since it is isolated and, located nearby at 8 parsec, it is spectroscopically significantly more accessible. From our $^{13}$CO detection in TYC 8988 b, it is expected that $^{13}$CO should be straightforward to see in 2 hrs of NIRSPEC data[41] from the KECK telescope, in particular at a spectral resolving power of $\lambda/\Delta\lambda$ ~27500. Our detailed analysis will be presented in a forthcoming publication. Extended Data Figure 6 shows the spectrum, best-fit model, and the $^{13}$CO cross-correlation signal, after performing an identical analysis - targeting the same features as our SINFONI data of TYC 8998 b. We retrieve a $^{12}$CO/$^{13}$CO isotopologue abundance of ~90-100, similar to the solar value. It will be intriguing to compare measurements between isolated brown dwarfs and exoplanets, revealing the role of carbon isotopologue ratio as a tracer of planet formation.

**Acknowledgements** We thank Ewine van Dishoeck, Alex Cridland and Anna Miotello for discussions on carbon fractionation in protoplanetary disks. We thank Katy Chubb for a $^{13}$CO line list comparison. Based on observations collected at the European Southern Observatory under ESO programme 2103.C-5012(C). Y.Z. and I.S. acknowledge funding from the European Research Council (ERC) under the European Union's Horizon 2020 research and innovation program under grant agreement No 694513. The research of A.J.B. and F.S. leading to these results has received funding from the European Research Council under ERC Starting Grant agreement 678194 (FALCONER). P.M. acknowledges support from the European Research Council under the European Union's Horizon 2020 research and innovation program under grant agreement No 832428. Part of this research was carried out at the Jet Propulsion Laboratory, California Institute of Technology, under a contract with the National Aeronautics and Space Administration.

**Author Contributions** Y.Z. and I.S. performed the data analysis and wrote the manuscript. A.J.B. led the SINFONI proposal, planned the observations and commented on the manuscript, and is the principal investigator (PI) of the Young Suns Exoplanet Survey (YSES) that led to the discovery of the TYC 8998 system. P.M. developed the retrieval models and assisted the data analysis. C.G, M.A.K., E.E.M, T.M., M.R., and F.S. constitute the core team YSES, contributed to the SINFONI proposal, and commented on the manuscript. H.J.H helped the preparations of the observations and commented on the manuscript.

**Competing interests** The authors declare no competing interests.

**Correspondence and requests for materials** should be addressed to I.S.

**Reprints and permissions information** is available at www.nature.com/reprints

**Data availability** The data is publicly available from the ESO Science Archive with the program ID: 2103.C-5012(C).

**Code availability** The data analysis was performed with custom Python scripts following the standard procedure. The code and reduced spectrum are available from https://gitlab.strw.leidenuniv.nl/yzhang/yses1b-sinfoni. The atmospheric retrieval models use the petitRADTRANS which is available from https://petitradtrans.readthedocs.io/, and the nested sampling tool PyMultiNest which is available from https://johannesbuchner.github.io/PyMultiNest/.

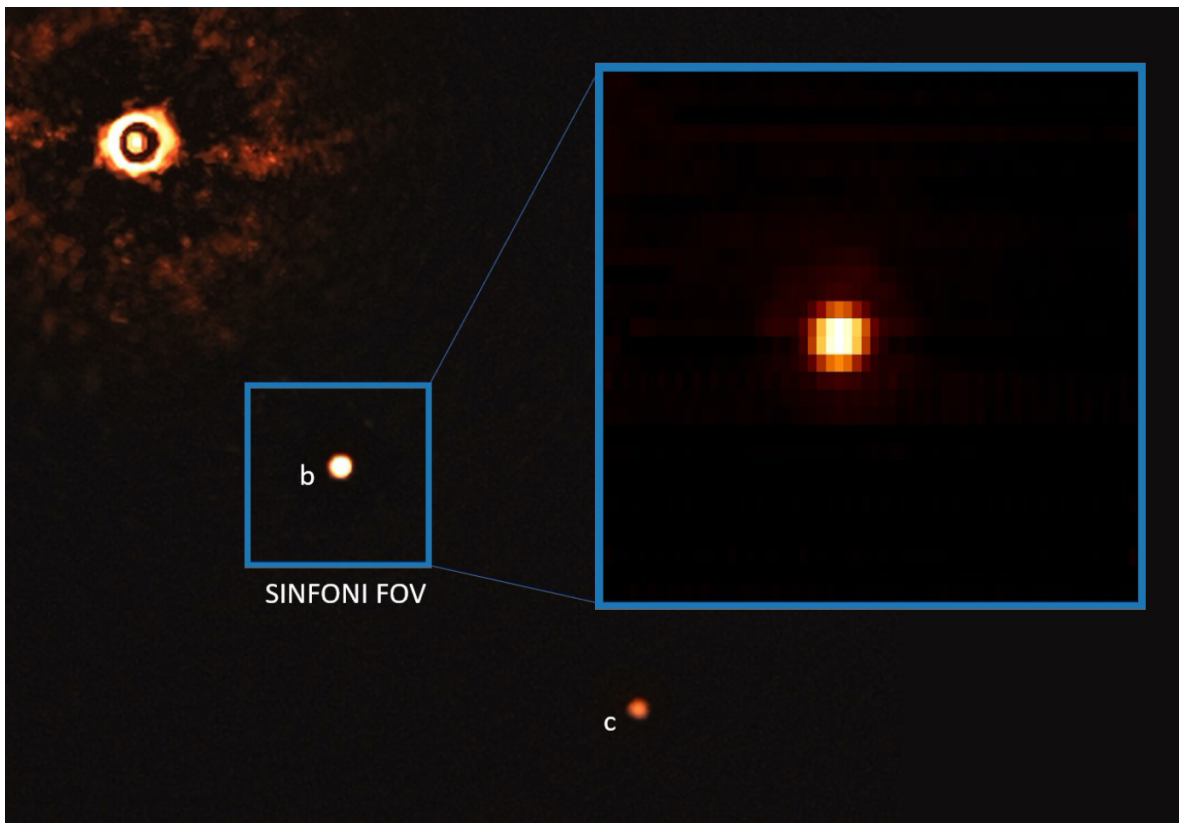

**Extended Data Figure 1 | Schematics of the observations of TYC 8998-760-1 b using SINFONI at the Very Large Telescope.** The background image is captured by the SPHERE instrument on the VLT (Credit: ESO/Bohn et al.). The small blue box marks the field of view (FOV) of SINFONI observations targeting the planet b. Both the host star and planet c are outside the FOV. An example of the wavelength-collapsed image is shown in the enlarged blue box, showing negligible contribution from starlight.

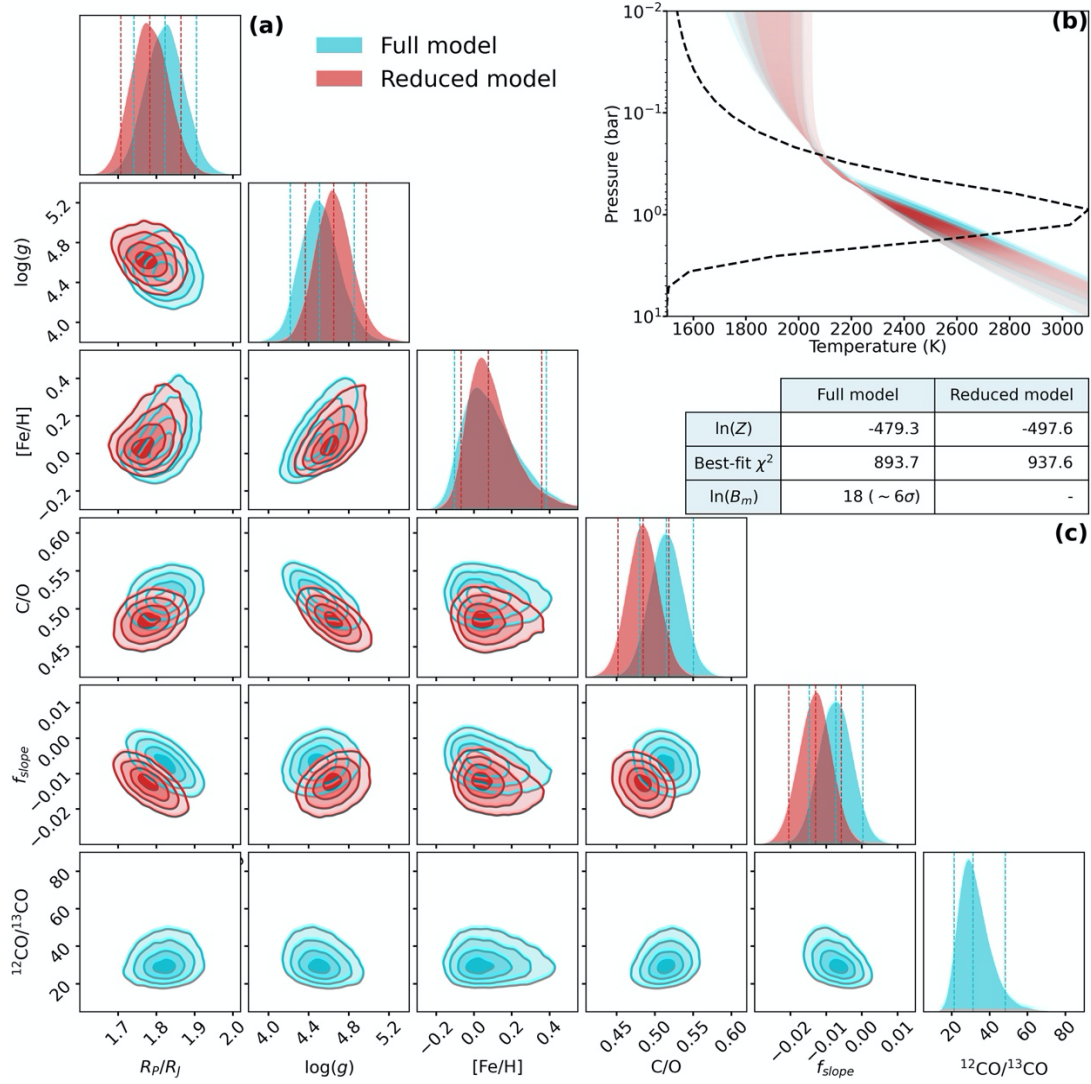

**Extended Data Figure 2 | Posteriors of the retrieved parameters and temperature structure for the full (cyan) and reduced (red) models.** The vertical dashed lines denote the 5%, 50%, 95% quantiles (90% uncertainties) of the distribution. Panel (b): T-P profile. The shaded regions with decreasing color saturation show 1σ, 2σ, and 3σ temperature uncertainty envelopes respectively. The black dashed line shows the flux average of the emission contribution function. The opaqueness of the temperature uncertainty envelopes has been scaled by this contribution function. Panel (c): fitting statistics of the full and reduced retrieval model, where ln(Z) and ln($B_m$) represent the logarithm of Bayesian evidence and Bayes factor.

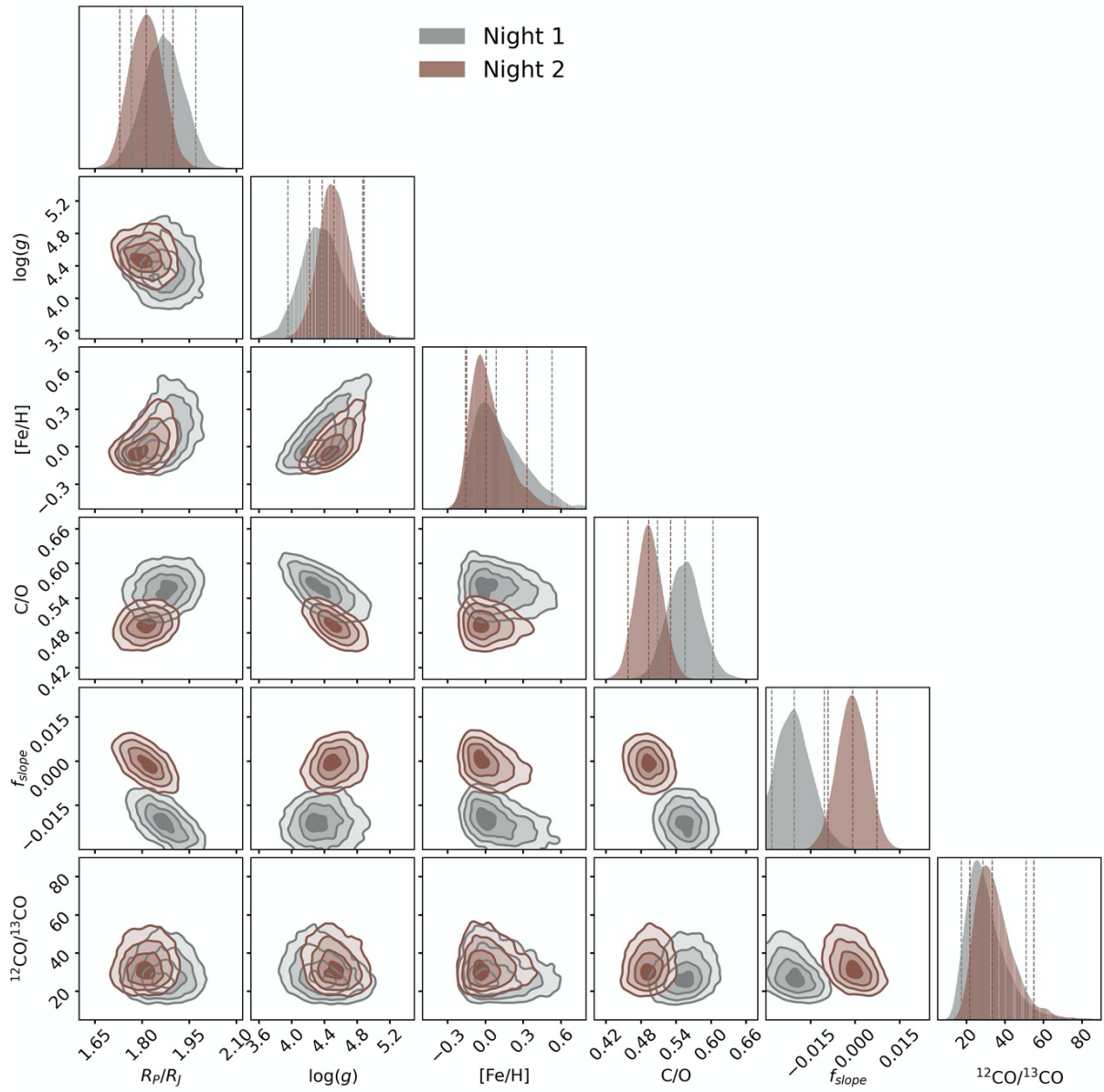

**Extended Data Figure 3 | Posteriors of the retrieved parameters for the data of individual nights.** Similar as Extended Data Figure 2a.

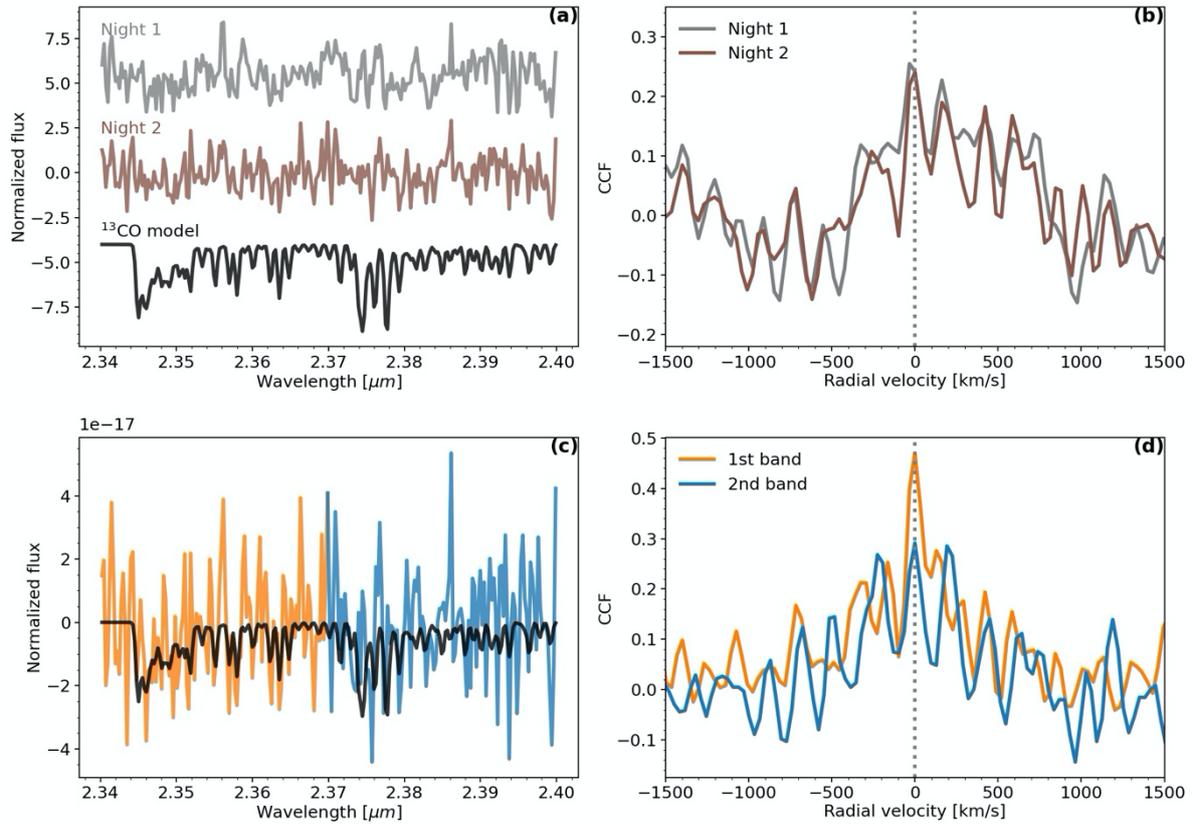

**Extended Data Figure 4 | Cross-correlation signal of $^{13}$CO from individual nights and bandheads.** Panel (a): observational residuals of two nights separately. Panel (b): cross-correlation signal from individual nights. Panel (c): filtered observational residuals of two $^{13}$CO bandheads separately. Panel (d): cross-correlation signal from individual bandheads.

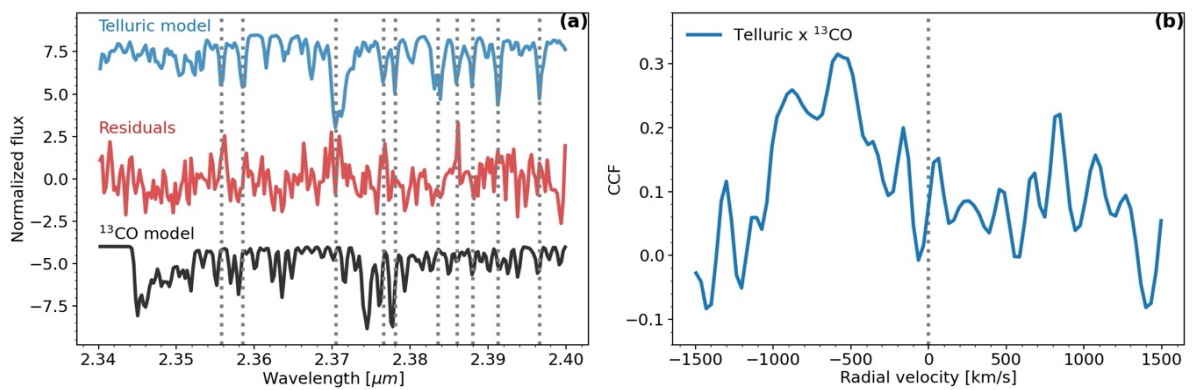

**Extended Data Figure 5 | Impact of telluric absorption lines and cross-correlation signal of $^{13}$CO at the extended wavelength region.** Panel (a): Comparison of the telluric transmission model with residuals. Some noise is attributed to imperfect telluric correction as noted by dotted gray lines. Panel (b): Cross-correlation function between the telluric model and the $^{13}$CO model, showing no correlation between them.

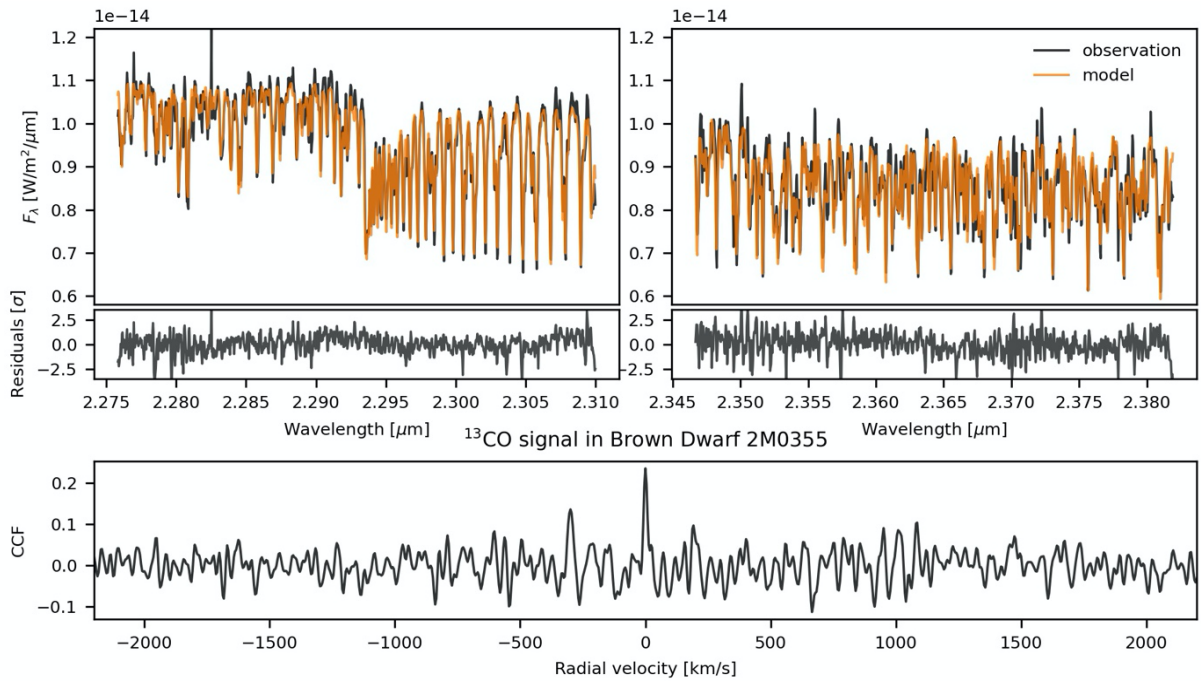

**Extended Data Figure 6 | K-band spectrum of the brown dwarf 2M0355 taken by Keck/NIRSPEC and the cross-correlation signal of $^{13}$CO.** The black line shows the observed spectrum and the orange line is the best-fit model obtained by retrieval analysis. Bottom panel: Cross-correlation function (CCF) between the $^{13}$CO model and observational residuals. The peak at zero velocity clearly shows the detection of $^{13}$CO.

| Parameter | Prior | Posterior (90%) |
|---|---|---|
| $R_P$ [$R_J$] | $U(1, 3)$ | $1.82 \pm 0.08$ |
| $\log(g)$ [$\log(\text{cm s}^{-2})$] | $U(2.5, 6)$ | $4.51^{+0.34}_{-0.29}$ |
| [Fe/H] | $U(-1.5, 1.5)$ | $0.07^{+0.31}_{-0.18}$ |
| C/O | $U(0.1, 1.5)$ | $0.52^{+0.04}_{-0.03}$ |
| $T_{int}$ [K] | $U(1000, 3000)$ | $2174^{+162}_{-150}$ |
| $\alpha$ | $U(0.1, 2)$ | $0.8^{+0.8}_{-0.2}$ |
| $\log(P_{phot})$ [bar] | $U(-3, 2)$ | $-0.19^{+0.10}_{-0.29}$ |
| $\log(^{13}CO/^{12}CO)$ | $U(-12, 0)$ | $-1.49^{+0.16}_{-0.19}$ |
| $f_{slope}$ | $U(-0.05, 0.05)$ | $-0.007 \pm 0.007$ |

**Extended Data Table 1 | Priors and inferred posteriors of the TYC 8998 b retrieval.** $U$(a, b) represents a uniform distribution ranging from a to b.

**Corresponding Author:**

Ignas Snellen
Leiden Observatory, Leiden University, Postbus 9513, 2300 RA Leiden, The Netherlands
snellen@strw.leidenuniv.nl
+31-71-5275838